\documentclass[aps, preprint, eqsecnum, showpacs, epsfig]{revtex4}
\usepackage{graphics}
\usepackage{graphicx}
\usepackage{epsfig}
\def \be {\begin{equation}}
\def \ee {\end{equation}}
\def \ba {\begin{eqnarray}}
\def \ea {\end{eqnarray}}
\def \bm {\begin{displaymath}}
\def \em {\end{displaymath}}
\def \br {{\bf r}}
\def \bom {{\bf \Omega}}

\begin{document}
\title{Effect of shape anisotropy on the phase diagram of the Gay-Berne fluid}
\author{Pankaj Mishra and Jokhan Ram}
\affiliation{Department of Physics, Banaras Hindu University, Varanasi-221 005, India}
\date{\today}
\begin{abstract}
We have used the density functional theory to study the effect of molecular elongation on 
the isotropic-nematic, isotropic-smectic A and nematic-smectic A phase transitions 
of a fluid of molecules interacting via the Gay-Berne intermolecular potential. 
We have considered a range of length-to-width parameter $3.0\leq x_0\leq 4.0$ in steps of 
0.2 at different densities and temperatures.  Pair correlation functions needed as input 
information in density functional theory are calculated using the Percus-Yevick 
integral equation theory. Within the small range of elongation, the  phase diagram 
shows significant changes. The fluid at low temperature is found to freeze directly
from isotropic to smectic A phase for all the values of $x_0$ considered by us on increasing
the density while nematic phase stabilizes in between isotropic and smectic A phases 
only at high temperatures and densities. Both isotropic-nematic and nematic-smectic A 
transition density and pressure are found to decrease as we increase $x_0$.  
The phase diagram obtained is compared with computer simulation result of the same 
model potential and is found to be in good qualitative agreement.
\end{abstract}
\maketitle
\section{\bf Introduction}
Liquid crystal phases which are formed by highly anisotropic complex organic molecules 
have symmetries intermediate between those of isotropic liquid and crystals. The simplest of the 
liquid crystal phases that find application in many electro-optic devices are nematic (N) and 
smectic A (Sm A) ones. In the nematic phase, molecules tend to align along a 
preferred direction called director, breaking the rotational invariance of 
isotropic liquid (I) but not the translational invariance. Partial breakdown of 
translational invariance along with breaking of
rotational invariance introduces smectic phases where molecules are essentially confined in
layers. In Sm A phase molecules are aligned perpendicular to layers with no intralayer or 
interlayer correlation in positions of the center of mass of the molecules. The properties 
and relative stability of these phases are extremely sensitive to the details of molecular 
structure and the true nature of intermolecular interaction potential[1]. Therefore, it is  
of interest for either computer simulation or theory to study the 
effect of molecular shape anisotropy and intermolecular potential on the liquid 
crystalline phase behaviour and properties.    

Because of the complex structure of the mesogenic molecules it is very difficult 
to know the exact nature of the interaction potential as a 
function of intermolecular separation and orientation.
Therefore, modeling of the intermolecular potential only with more physically 
relevant features become inevitable. The pair interaction potential 
model proposed by Gay and Berne (GB)[2]
is one such model that includes anisotropic attractive interactions along with short range 
repulsive interactions. It has now become a standard model to study liquid 
crystalline phases and has been widely used in the investigation of various phenomena
through computer simulation[3-13] and also theoretically[14-18].          

In the Gay-Berne (GB) pair potential model, the molecules are considered as ellipsoids
of revolution about the principal axis of the molecule. The interaction potential between two
such ellipsoidal molecules $i$ and $j$ depends on direction $\bf \hat r_{ij}$ and on the 
magnitude of 
center-center vector $\bf {r_{ij}= r_{i}-r_{j}}$ and upon molecular axis vectors 
$\bf{\hat e_i}$ and $\bf{\hat e_j}$. 
\newpage
The GB potential is expressed as
\ba
u({\hat {\bf e_i}}, {\hat {\bf e_j}}, {\bf r_{ij}})=
&& 4\epsilon_{0}\epsilon^{\nu}({\hat {\bf e_i}}, 
{\hat {\bf e_j}}){\epsilon'}^{\mu} ({\hat {\bf e_i}}, 
{\hat {\bf e_j}}, {\hat\br_{ij}})\times \nonumber \\
& &\left[\left(\frac{r_{ij}-\sigma({\hat{\bf e_i}}, 
{\hat{\bf e_j}}, {\hat\br_{ij}}) + \sigma_0}{\sigma_0}\right)^{-12}
\right. \nonumber \\
& & -\left. \left(\frac{r_{ij}-\sigma({\hat{\bf e_i}}, 
{\hat{\bf e_j}},{\hat\br_{ij}})+\sigma_0}{\sigma_0}\right)^{-6}\right]
\ea
The angle dependent range parameters $\sigma$
and strength functions $\epsilon$ are given by
\ba
\sigma({\hat{\bf e_i}}, {\hat{\bf e_j}}, {\hat{\bf r_{ij}}})=
&&\sigma_0 \left[1-\chi\left(\frac{({\hat{\bf e_i}}.{\hat\br_{ij}})^2
+ ({\hat{\bf e_j}}.{\hat\br_{ij}})^2}
{1 - \chi^{2}({\hat{\bf e_i}}.{\hat{\bf e_j}})^2}
\right. \right. \nonumber \\
& & -\left. \left. \frac{2\chi({\hat{\bf e_i}}.{\hat\br_{ij}})({\hat{\bf e_j}}.{\hat\br_{ij}})
({\hat{\bf e_i}}.{\hat{\bf e_j}})}
{1 - \chi^{2}({\hat{\bf e_i}}.{\hat{\bf e_j}})^2} \right) \right]^{-\frac{1}{2}}
\ea

\ba
\epsilon ({\hat {\bf e_i}}, {\hat {\bf e_j}}) = [1 - \chi^2
({\hat {\bf e_i}}.{\hat {\bf e_j}})^2]^{-\frac{1}{2}}
\ea

\ba
\epsilon'({\hat {\bf e_i}}, {\hat {\bf e_j}}, {\hat\br_{ij}})= 
&& \left[1-\chi'\left(\frac{({\hat{\bf e_i}}.{\hat\br_{ij}})^2
+ ({\hat{\bf e_j}}.{\hat\br_{ij}})^2}
{1 - \chi^{\prime 2}({\hat{\bf e_i}}.{\hat{\bf e_j}})^2} 
\right. \right. \nonumber \\ 
& & -\left. \left. \frac{2\chi'({\hat{\bf e_i}}.{\hat\br_{ij}})({\hat{\bf e_j}}.{\hat\br_{ij}})
({\hat{\bf e_i}}.{\hat{\bf e_j}})}
{1 - \chi^{\prime 2}({\hat{\bf e_i}}.{\hat{\bf e_j}})^2} \right) \right]
\ea

where $\sigma_0$ is the smallest molecular diameter and $\epsilon_0$ is the 
energy scaling parameter equal to the well depth for the cross configuration
$(\hat{\bf e_i}.{\hat{\bf e_j}}=\hat\br_{ij}.\hat{\bf e_i}=\hat\br_{ij}.\hat{\bf e_j}=0)$.
The parameter $\chi$ is a measure of shape anisotropy defined as 
$\chi=(x_{0}^2-1)/(x_{0}^2+1)$ with
$x_0$ being the length-to-width ratio of the molecule.Though $x_0$ measures the anisotropy 
of the repulsive core, it also determines the difference in the depth of the attractive 
well between the side-by-side and the cross configurations. 
The parameter $\chi'$ determines the energy anisotropy defined as 
$\chi' = (k'^{ 1/\mu}-1)/(k'^{ 1/\mu}+1)$, 
where $k'$ is the well-depth ratio for side-by-side and end-to-end configuration.
The powers $\mu$ and $\nu$ entering in the energy functions are adjustable 
exponents determining the strength of interaction.  

The GB model contains four parameters $(x_0, k', \mu, \nu)$ that determine 
the anisotropy in the repulsive and attractive forces in addition to two 
parameters $(\sigma_0, \epsilon_0)$ that scale the distance and energy, respectively.
The choices for the values of these parameters are in no way unique and they can be varied
to yield a wide range of anisotropic potential[3]. The most commonly used values of
these parameters in the literature are $x_0=3.0, k'=5, \mu=2, \nu=1$. 
In their molecular dynamic simulation study for this set of parameters 
de Miguel $\it {et} {al}$[4, 5] found three phases, namely,
isotropic, nematic and smectic B. No Sm A phase was found to be stable. However, existence 
of Sm A phase has been reported for the set of parameters $x_0=3.0, k'=5, \mu=1, \nu=2$[6].
Recently the effect that changes in the well depth parameter $k'$[7, 15] and molecular elongation 
$x_0$[8, 9, 17] have upon the overall phase behaviour of the system has been investigated. 
It was found that with the increase in $k'$ 
the Sm B phase is favored at low density while the nematic phase becomes increasingly stable 
at lower temperature as $k'$ is decreased. The variation of molecular elongation parameter
$x_0$ has been found to have a significant effect. 
An island of Sm A is found to appear in the phase diagram for elongations above 
$x_0=3.0$[9]. The range of Sm A extends 
to both higher and lower temperatures as $x_0$ is increased. Also the
isotropic-nematic (I-N)transition is seen to move to lower density (and pressure) at a
given temperature with the increase in $x_0$.           

In this paper we use the density functional theory (DFT) to study the effect of 
the variation of the length-to-breadth ratio $x_0$ on the freezing transitions and 
freezing parameters for the I-N, I-Sm A and N-Sm A transition for a system of molecule
interacting via the Gay-Berne pair potential. The value of $x_0$ has been varied from 
3.0 to 4.0 in steps of 0.2, keeping other parameters fixed at $k^{'}=5, \mu=2$ and $\nu=1$. 
Pair correlation functions needed as input information in the DFT have been calculated using 
the Percus-Yevick (PY) integral equation theory. The paper is organized as follows: 
In section II we describe in brief the density-
functional formalism used to locate the freezing transitions and freezing parameters.
We discuss our results and compare with the available simulation results in section III and
finally conclude our discussion in section IV.  

\section{\bf Fundamentals of the density-functional theory of freezing}
The density functional theory (DFT) directly links the bulk phase behaviour 
of a fluid with its molecular properties.
In DFT, the equilibrium density profile of a non uniform anisotropic liquid can 
be determined as the one that minimizes the
grand thermodynamic potential $W$ regarded as a functional of the single-particle 
density function $\rho(\br,\bom)$ at point $\br$ and orientation $\bom$.

The grand thermodynamic potential has the general form
\begin{equation}
-W = \beta A - \beta {\mu_c} \int d{\bf x} \rho({\bf x})
\end{equation}
where A is the Helmholtz free energy, $\mu_c$ the chemical potential and
$\rho({\bf x})$ is a singlet distribution function, to locate the transition.

The above mentioned minimum property of the grand thermodynamic potential follows from
the variational inequality
\be
W[\rho_{eq}] \le W[\rho]
\ee
which is valid for a fixed temperature and chemical potential, where $\rho_{eq}$ is the 
equilibrium density profile.

It is convenient
to subtract the isotropic fluid thermodynamic potential from $W$ and write it as [19]
\begin{equation}
\Delta W = W - W_f = \Delta W_1 + \Delta W_2
\end{equation}
with
\begin{eqnarray}
\frac{\Delta W_1}{N} & = & \frac{1}{\rho_f V} \int {d\bf r}
{d\bf \Omega}\left\{\rho({\bf r}, {\bf \Omega})
\ln \left[\frac{\rho({\bf r}, {\bf \Omega})}{\rho_f}\right] 
\right. \nonumber \\ 
&&\left. -\Delta \rho({\bf r}, {\bf \Omega})\right\} \\
{\rm and} \nonumber \\
\frac{\Delta W_2}{N} & = & -\frac{1}{2\rho_f} \int {d\bf r_{12}}
{d\bf \Omega_1}{d\bf \Omega_2}\Delta\rho({\bf r_1}, {\bf \Omega_1}) \times \nonumber \\
& &c({\bf r_{12}}, {\bf \Omega_1}, {\bf \Omega_2})\Delta\rho({\bf r_2},
{\bf \Omega_2})
\end{eqnarray}
Here $\Delta\rho({\br, \bom}) = \rho({\br,\bom }) - \rho_f$, where
$\rho_f$ is the density of the coexisting liquid.

The order parameter equation is obtained by minimizing $\Delta W$ with respect
to the arbitrary variation in the ordered phase density subject to a constraint that 
corresponds to some specific feature of the ordered phase. This leads to
\be
\ln \frac{\rho(\br_1, \bom_1)}{\rho_f}=\lambda_L+\int d\br_2 d\bom_2 
c(\br_{12},\bom_1,\bom_2;\rho_f)\Delta\rho(\br_2, \bom_2)
\ee
where $\lambda_L$ is Lagrange multiplier.
Equation(10) is solved by expanding the singlet distribution $\rho({\br, \bom})$ in
terms of the order parameters that characterize the ordered structures using
the Fourier series and Wigner rotation matrices . Thus
\be\
\rho(\br, \bom) = \rho_0 \sum_q \sum_{lmn} Q_{lmn}
(G_q) \exp(i{\bf G}_q.\br) D^l_{mn}(\bom)
\ee\
where the expansion coefficients
\ba\
 Q_{lmn}(G_q) &=& \frac{2l+1}{N}\int d\br \int d\bom \rho(\br, \bom)\times\nonumber\\
&&\exp(-i{\bf G_q}.\br) D^{*l}_{mn}(\bom)
\ea\
are the order parameters, ${\bf G_q}$ the reciprocal lattice vectors, $\rho_0$
the mean number density of the ordered phase and $D^{*l}_{mn}(\bom)$ the generalized spherical
harmonics or Wigner rotation matrices. Note that for a uniaxial system consisting
of cylindrically symmetric molecules $m=n=0$ and, therefore, one has
\be\
\rho(\br, \bom) = \rho_0 \sum_l \sum_q Q_{lq} \exp(i{\bf G}_q.\br)
                  P_l(\cos \theta)
\ee\
and
\be\
Q_{lq} = \frac{2l+1}{N}\int d\br \int d\bom \rho(\br, \bom)
\exp(-i{\bf G_q}.\br) P_l(\cos \theta)
\ee\
where $ P_l(\cos \theta)$ is the Legendre polynomial of degree $l$ and
$\theta$ is the angle between the cylindrical axis of a molecule and the
director.

In the present calculation we consider two orientational order parameters
\ba
\bar P_l&=&\frac{Q_{l0}}{2l+1}=\langle P_l(cos\theta)\rangle
\ea
with $l$=2 and 4, one order parameter corresponding to positional order
along Z axis,
\ba
{\bar\mu}&=&Q_{00}(G_z)=\langle cos(\frac{2\pi}{d}z)\rangle
\ea
($d$, being the layer spacing) and one mixed order parameter that measures the
coupling between the positional and orientational ordering and is defined as,
\ba
\tau&=&\frac{1}{5}Q_{20}(G_z)=\langle cos(\frac{2\pi}{d}z) P_{l}(cos\theta)\rangle
\ea
The angular bracket in above equations indicate the ensemble average.

The following order parameter equations are obtained by using Eqs.(13) and (17) [18];
\ba
\bar P_l &=&\frac{1}{2d}\int_{0}^{d}dz_{1}\int_{0}^{\pi}sin\theta_1 d\theta_1
P_l(cos\theta_1)\exp[sum]\\
{\bar\mu} &=&
 \frac{1}{2d}\int_{0}^{d}dz_{1}\int_{0}^{\pi}sin\theta_1 d\theta_1
cos(\frac {2\pi z_1}{d})\exp[sum]\\
\tau &=&
\frac{1}{2d}\int_{0}^{d}dz_{1}
\int_{0}^{\pi}sin\theta_1 d\theta_1
P_2(cos\theta_1)cos(\frac{2\pi z_1}{d}) \times \nonumber \\
& & \exp[sum]
\ea
and change in density at the transition is found from the relation
\be
1+\Delta\rho^{*}=\frac{1}{2d}\int_{0}^{d}dz_{1}\int_{0}^{\pi}sin\theta_1 d\theta_1
\exp[sum].
\ee
Here
\ba
sum &=& \Delta\rho^{*}\hat C_{00}^0+2{\bar\mu} cos(\frac{2\pi z_1}{d})\hat C_{00}^1(\theta_1)+
\bar P_2 \hat C_{20}^{0}(\theta_1)+ \nonumber \\
&& \bar P_4 \hat C_{40}^{0}(\theta_1)+
2\tau cos(\frac{2\pi z_1}{d}) \hat C_{20}^{1}(\theta_1)
\ea
and
\ba
\hat C_{L0}^{q}(\theta_1)&=&({\frac{2l+1}{4\pi}})^{1/2}
\rho_f\sum_{l_1l}i^l(2l_1+1)^{1/2}(2l+1)^{1/2}\times \nonumber \\
&&P_{l_{1}}(cos\theta_1)C_g(l_1Ll;000)\times \nonumber \\
&&\int_{0}^{\infty}c_{l_1Ll}(r_{12}) j_{l}(G_{q}r_{12})r_{12}^2 dr_{12}
\ea
where $c_{l_1Ll}(r)$ are direct pair correlation function harmonics 
,$C_{g}(l_1Ll;000)$ are Clebsch-Gordan coefficients and $G_q=2\pi/d$.

\begin{figure}[h]
\includegraphics[height=3.0in, width=3.5in]{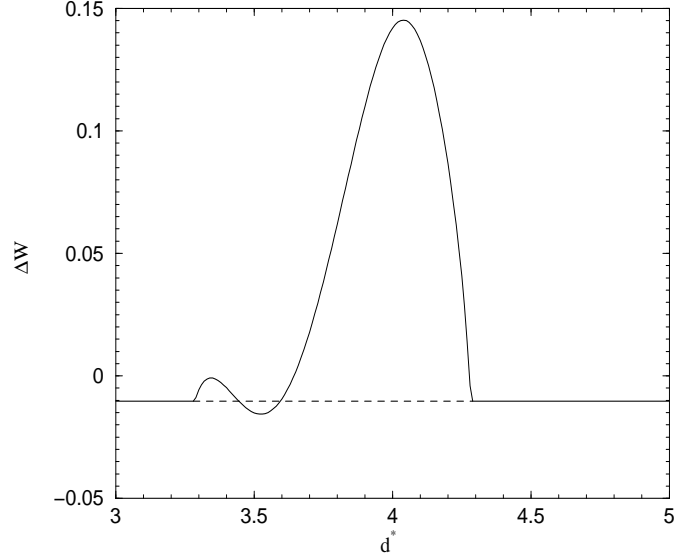}
\caption{Variation of grand thermodynamic potential with smectic interlayer spacing
for nematic-smectic A transition density $\rho^*=0.219$ at $T^*=1.40$ for the 
GB potential parameters $x_0= 4.0, k'=5, \mu=2$ and $\nu=1$.}
\end{figure}

In order to evaluate the transition parameters such as order parameters, change in density
etc, equation (15)-(21) were solved self consistently using values of the 
harmonics $c_{l_1Ll}(r)$ evaluated at given temperature and density for each value of
$x_0$. The calculation is repeated with different values of Sm A interlayer spacing, 
$d$. By substituting these solutions in Equations (7)-(9) we find the grand thermodynamic 
potential difference between ordered and isotropic phases; i.e.
\ba
-\frac{\Delta W}{N}&=& -\Delta\rho^{*}+\frac{1}{2}\Delta\rho^{*}(2+\Delta\rho^{*})\hat C_{00}^{0}+
\frac{1}{2}(\bar P_{2}^{2}\hat C_{22}^{0}+\nonumber \\
&&\bar P_{4}^{2}\hat C_{44}^{0})+
 {\bar\mu}^{2}\hat C_{00}^{1}+
2{\bar\mu}\tau\hat C_{20}^{1}+\tau^{2}\hat C_{22}^{1}
\ea
where
\ba
\hat C_{LL'}^{q}&=&(2L+1)^{1/2}(2L'+1)^{1/2}\times \nonumber \\
&& \rho_f\sum_l i^l({\frac{2l+1}{4\pi}})^{1/2}
C_g(LL'l;000)\times\nonumber\\
&&\int_{0}^{\infty} c_{LL'l}(r_{12}) j_{l}({ G_q r_{12}}) r_{12}^{2} dr_{12}
\ea

At given temperature and density the phase with lowest grand potential is taken as
the stable phase. Phase coexistence occurs at the value of $\rho_f$ that makes
${-\Delta W}/{N}=0$ for the ordered and the liquid phases. The transition
from nematic to the SmA is determined by comparing the values of $-{\Delta W}/{N}$
of these two phases at a given temperature and at different densities.
The value of the interlayer spacing $d$, is found by minimizing the grand
potential with respect to d. In Fig.1 we plot the variation of $\Delta W$ 
with $d^*(=d/\sigma_0)$. After selecting the value of $d$ for a given
density and temperature we locate the transition point using the procedure
outlined above. In the isotropic phase all the four order parameters become zero. In the nematic 
phase the orientational order parameters $\bar P_2$ and $\bar P_4$ become
nonzero but the other two parameters $\bar\mu$ and $\tau$ remain zero.
This is because the nematic phase has no long range
positional order. In the SmA phase all the four order parameters are
nonzero showing that the system has both long range orientational
and positional order along one direction.

\section{\bf Results and Discussion}
The pair correlation functions(PCF's) of molecular fluids are the lowest order
microscopic quantities which contain information about the structure of
the fluid as well as have direct contact with the underlying intermolecular
interactions. Most of the structural informations of an ordered phase are
contained in single particle distribution $\rho(x)$, as shown in the previous 
section. In the DFT of freezing the single particle distribution $\rho(x)$
of an ordered phase is expressed in terms of the PCF's 
of the isotropic fluid. In this study, the PCF's of the isotropic phase 
are found by solving Ornstein-Zernike (OZ) equation using Percus-Yevick (PY)
closure (details are same as in reference [18]) at different values of the reduced 
temperatures $T^*$, in the range from 0.80 to 1.80 and at various densities for
each $3.0\le x_0 \le 4.0$. 
\begin{figure*}
\includegraphics[width=4.0in, angle=0]{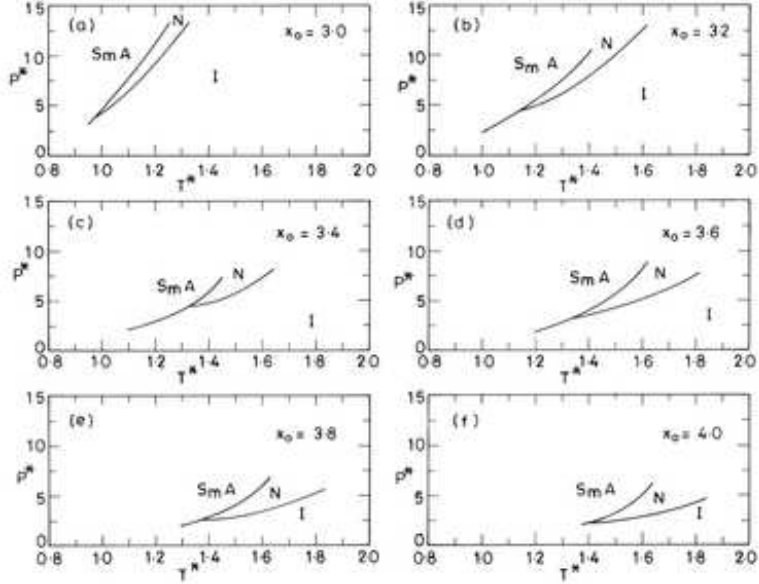}
\caption{Pressure-Temperature phase diagrams for the GB potential with parameters
$3.0 \le x_0\le 4.0, k'=5, \mu=2$ and $\nu=1$ using density-functional theory}
\end{figure*}

\begin{figure}[h]
\includegraphics[height=3.0in, width=3.5in]{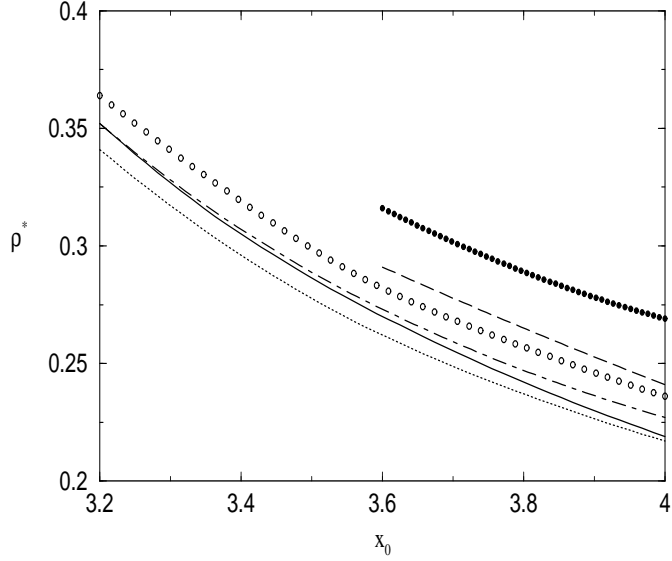}
\caption{Variation of I-N and N-Sm A transition densities with molecular elongation
parameter $x_0$. Dotted and solid lines represent, respectively,
 I-N and N-Sm A transition densities at $T^*=1.40$. Dot-dashed and long dashed 
lines are the respective lines at $T^*=1.50$ and
open circles and filled circles are those at $T^*=1.60$, respectively.} 
\end{figure}

In Figures 2(a-f) we present the phase diagrams in the pressure-temperature 
plane. For all $x_0\ge 3.0$, we first find the direct I-Sm A transition below a certain
temperature $T^*$ which increases from $T^*\simeq0.99$ for $x_0=3.0$
to $T^*\simeq1.39$ for $x_0=4.0$. The lowest temperature where Sm A phase starts stabilizing
is found to rise as we increase $x_0$. On further increase in $T^*$, nematic phase enters
into the phase sequence. Temperature range of nematic stability is found to 
increase as molecular elongation is increased which is also reflected in the ratio
of N-Sm A and I-N transition temperatures, $T_{N-A}/T_{I-N}$. For example, the 
value of the ratio change from 0.96 to 0.92 as we move from $x_0= 3.6$ to $x_0= 4.0$
with pressure, $P^*$, fixed at 3.5. This corresponds to a higher orientational order
before the Sm A phase stabilizes. According to McMillan theory [20], this value of the 
$T_{N-A}/T_{I-N}$ ratio predicts that the N- Sm A transition is first order. Figs. 2(a-f) also
show that as $x_0$ is increased, both the I-N and N-Sm A transitions move to a lower
pressure.   
\begin{figure}[h]
\includegraphics[height=3.0in, width=3.5in]{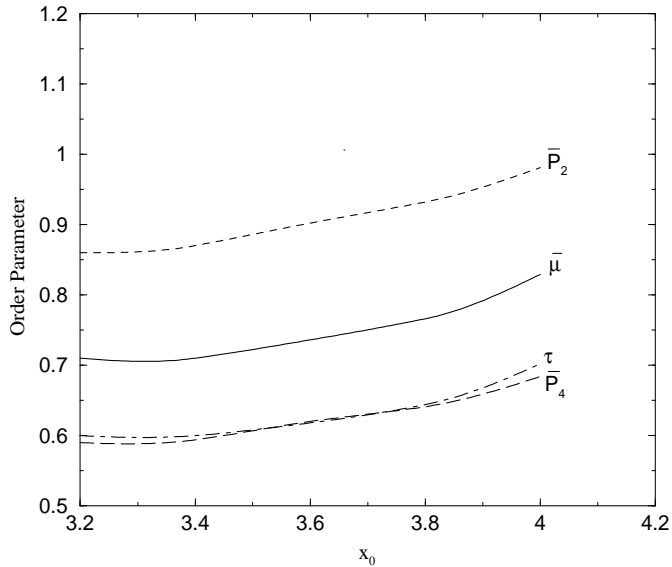}
\caption{Variation of order parameters with $x_0$ for N-Sm A transition 
at $T^*=1.40$ keeping other GB potential parameters fixed at 
 $k'=5, \mu=2$ and $\nu=1$.}
\end{figure}

In Figures 3, we plot the I-N and N-Sm A transition densities against the molecular
elongation $x_0$. Both the I-N and N-Sm A transition densities are seen to decrease
with the increase in $x_0$ in agreement to reference [9]. It can be seen from this figure
that the difference between the I-N and N-Sm A transition densities decreases as 
$x_0$ is increased. This is more apparent in the low temperature curves where the
region between the two transition lines gradually shrinks. This makes evident that
for a given temperature, with the increase in molecular elongation, the Sm A phase become
more  probable at low density and eventually takes over nematic resulting in direct
isotropic to Sm A transition. This stabilization of smectic phase is to be expected since
increase in $x_0$ results in deeper well depth for parallel configuration of molecules
making it energetically more favourable. Also the I-N and N-Sm A transition density difference 
is seen to decrease rapidly as the temperature is reduced at a fixed $x_0$, showing the effect
of the attractive intermolecular interaction which dominates at low temperature facilitating
the formation of the Sm A Phase. This also indicates the possibility of direct I- Sm A transition
for a given $x_0$ below certain temperature where Sm A phase takes over nematic.            

The Sm A phase is characterized by the presence of a non zero translational and 
mixed order parameters along with the orientational order parameters. Fig.4 presents
the variation of the order parameters at N-Sm A transition with respect to $x_0$ at 
temperature (arbitrarily chosen) $T^*= 1.40$. It shows that the order parameters are
monotonically increasing with the increase in the molecular elongation.       
For each $x_0$, we found that the values of translational
and mixed order parameters, which are characteristics of Sm A phase, decrease
rapidly with the increase in temperature, eventually becoming zero. This shows that
the stable Sm A phase becomes progressively less ordered with the increasing temperature
and finally destabilizes with respect to nematic phase and disappears from the phase
diagram. This is obvious as attractive forces between the molecules are less 
important at high temperatures and purely repulsive intermolecular interactions
do not exhibit Sm A. One should note that at $x_0=3.0$, we get stable Sm A phase which 
has not been observed in computer simulation experiments. The phase observed in 
simulation [9] is Sm B which we have not considered in the present study. 
In Table 1 we have compared our results for the I-N transition parameters 
with the simulation results of refs.[12, 13] at temperature $T^*=1.00$ for 
$x_0=3.0$. It can be seen that the values of the transition parameters found 
by us are higher than those of the simulations but relatively close to the values of ref.[13].

\begin{table}[h]
\caption{Comparison of isotropic-nematic transition parameters of the GB(3, 5, 2, 1) 
potential at $T^*= 1.00$. Quantities in  reduced units are 
Pressure $P^* = P\sigma_0^3/\epsilon_0,$ and $ \mu^{*}_c = \mu_{c}/\epsilon_0$.}
\begin{tabular}{|c|c|c|c|c|c|c|} \hline
Theory&$\rho^*$ & $\Delta \rho^*$ &${\bar P_2}$ & ${\bar P_4}$ &P*&$\mu^{*}_{c}$ \\  \hline
    DFT& 0.336 & 0.017 & 0.675 & 0.373 & 4.41 & 14.57 \\
 MC[13]& 0.320 &   -   & 0.660 & 0.290 &  -   &  -    \\
 MC[12]& 0.307 & 0.017 & 0.520 &   -   & 3.63 & 12.79 \\ \hline
\end{tabular}
\end{table}

 The change in density at transitions 
$\Delta\rho^*(=\frac{\rho_n-\rho_l}{\rho_l})$ is found to increase as 
$x_0$ is increased at a given temperature. For example, for $T^*=1.50$, at I-N transition 
$\Delta \rho^*$ changes from $4.9\%$ to $6.0\%$ as $x_0$ is changed from 3.8 to 4.0, whereas
the corresponding change at the N-Sm A transition is from $3.4\%$
 to $4.7\%$. In the simulation [9], though the value of $\Delta\rho^*$ at the N-Sm A
transition increases from $\sim 1.6\%$ to $\sim 3.1\%$ as $x_0$ changes from
3.8 to 4.0, the trend at I-N transition is reverse where it decreases from 
$\sim8\%$ to $\sim5.2\%$.            

\begin{figure}[h]
\includegraphics[angle=0]{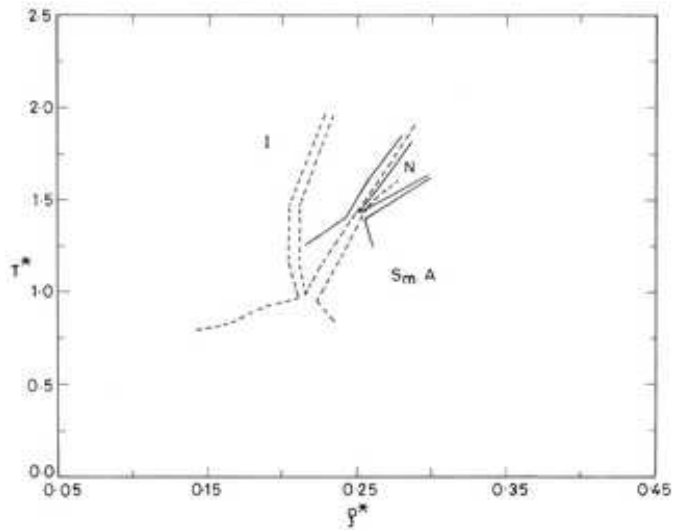}
\caption{Temperature-Density phase diagram for the GB potential with parameters 
$x_0=3.8, k'=5, \mu=2$
and $\nu=1$. The solid lines indicate the phase boundaries obtained by using the 
density-functional theory while dashed lines are the simulation results of Brown 
$\it et al $[9].}
\end{figure}
\begin{figure}[h]
\includegraphics[angle=0]{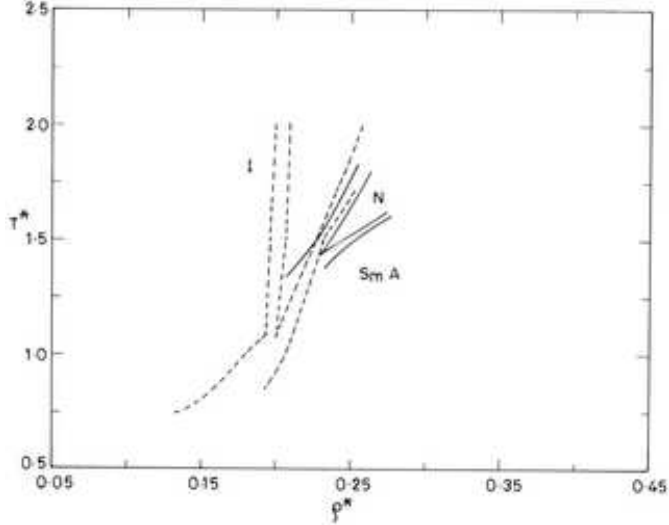}
\caption{Temperature-density phase diagram for the GB potential with parameters 
$x_0=4.0, k'=5, \mu=2$ and $\nu=1$. The curves are the same as in figure 5.}
\end{figure}

In Figs. 5 and 6 we present our phase diagram in the temperature-density plane
for $x_0=3.8$ and 4.0. The solid lines represent the phase boundaries  
calculated by us while dashed lines are the approximate phase boundaries taken from 
ref. [9] for the sake of comparison. Though the quantitative agreement 
between the phase diagrams shown in the Figures 5 and 6 is not so encouraging,
however, they agree qualitatively well. This may be due to 
the fact that in our calculation the transition takes place at higher density
than those found in the simulations. The critical correlations at which
the isotropic phase become unstable are found to be higher than actual values
because PY theory is known to underestimate the orientational correlations[21].
This shortcoming of the PY theory is found to increase with 
increasing temperature which explains why in Figs. 5 and 6 the I-N
transition boundaries are more shifted than the N-Sm A boundaries.     
\section{\bf Conclusion}
We have used the Gay-Berne potential model, to study the
effect of the variation of elongation parameter $x_0$
on its phase behaviour with the values of all other parameters 
fixed at $k'=5, \mu=2$ and $\nu=1$. The pair correlation functions of the 
isotropic fluid are calculated using Percus-Yevick integral equation theory
and have been used in density-functional theory to locate the 
isotropic-nematic, isotropic-smectic A and nematic-smectic A transitions.
Within the small range of elongation $3.0\le x_0 \le4.0$, the phase diagram
shows significant changes. We have discussed how various phase boundaries,
transition densities, pressure, change in density at transition, order parameters
etc. change with the variation in molecular elongation. We found a stable Sm A phase
for all $x_0\ge 3.0$. Both I-N and N-Sm A transitions are found to move to lower
density and pressure as $x_0$ is increased. We have compared our results 
with those of computer simulations
and found that the density-functional theory reproduces all the features of the 
phase diagrams which are in good qualitative agreement.
To have a better quantitative agreement there is a need to evaluate the 
isotropic pair correlations more accurately than those given by Percus-Yevick 
theory. Also the use of correlations evaluated 
directly in the nematic phase may improve the results. The work in this
direction is in progress.      
\section{\bf Acknowledgment}
We are grateful to Prof. Y. Singh for many helpful discussions and encouragement. The 
work was supported by Department of Science and Technology (India) through a project
grant.

\end{document}